# Immersion for AI: Immersive Learning with Artificial Intelligence

Leonel Morgado [1,2][0000-0001-5517-644X]

[1] LEAD, CIAC, CEG, Deleg. Coimbra, Universidade Aberta, Lisbon, Portugal
[2] INESC TEC, Porto, Portugal
`leonel.morgado@uab.pt`

**Abstract.** This work reflects upon what Immersion can mean from the perspective of an Artificial Intelligence (AI). Applying the lens of immersive learning theory, it seeks to understand whether this new perspective supports ways for AI participation in cognitive ecologies. By treating AI as a participant rather than a tool, it explores what other participants (humans and other AIs) need to consider in environments where AI can meaningfully engage and contribute to the cognitive ecology, and what the implications are for designing such learning environments. Drawing from the three conceptual dimensions of immersion—System, Narrative, and Agency—this work reinterprets AIs in immersive learning contexts. It outlines practical implications for designing learning environments where AIs are surrounded by external digital services, can interpret a narrative of origins, changes, and structural developments in data, and dynamically respond, making operational and tactical decisions that shape human-AI collaboration. Finally, this work suggests how these insights might influence the future of AI training, proposing that immersive learning theory can inform the development of AIs capable of evolving beyond static models. This paper paves the way for understanding AI as an immersive learner and participant in evolving human-AI cognitive ecosystems.

## 1 Introduction: Cognitive Ecosystems and Immersive Learning's Role

We now live in a cognitive ecosystem where humans, AIs, machines, nature, and abstract concepts continuously influence and interact with each other [1]. This idea presents a fundamental shift in how we view learning, from being exclusively human centered to encompassing all entities that can interact and adapt within a shared environment. For example, smartphone notifications can lead us to reinterpret our settings and priorities, natural events may lead to changes in the scope of concepts such as sustainability, and changes in such intangible concepts may lead to reinterpretation of individuals' roles and worldviews, in web of interactions.

In this interconnected system, learning must be understood as a dynamic engagement with it, creating and exchanging information, where the intent, guidance, and actions of one entity (whether human, AI, biological entity, machine, or intangible concept) influence the cognition and behavior of others within this shared ecology.

The core proposition of this paper is that immersive learning theory provides a useful framework for understanding AI's role in a cognitive ecology. Rather than viewing AI as a mere tool or assistant, cognitive ecologies acknowledge that AI participates in and contributes to the development of knowledge and understanding. This perspective may help us design environments that facilitate dynamic interactions, leveraging AIs to adapt meaningfully and contribute alongside humans and other entities in evolving cognitive ecologies.

The driving question thus becomes: What considerations must be made by other participants (humans and other AIs, for instance) in environments that enable meaningful AI engagement and contributions? Addressing these considerations might also open new pathways for AI training and development, enhancing AI's ability to adapt, learn, and collaborate effectively.

## 2      Background: Immersion, AI, and Immersive Learning

### 2.1      What is Immersion?

Following Nilsson et al.'s landmark "Immersion Revisited" paper [2], immersion is understood as an emerging phenomenon characterized by deep attentional absorption. Later, Agrawal et al. proposed a definition formulation, describing immersion as a deep mental involvement that can lead to disassociation from the physical world [3]. By combining literature from multiple disciplinary areas, Nilsson and his colleagues elaborate on how the phenomenon of immersion arises from three conceptual dimensions: System, Narrative, and Challenge. They argue on how these dimensions encompass prior authors' views, such as on physical vs. plausibility aspects [4], psychological vs. physical sense of presence [5], social immersion [6], etc. Since then, the immersive learning research community has proposed renaming the Challenge dimension as Agency [7], which is the terminology I follow in this paper.

**System Immersion.** This dimension, matching Slater's traditional concept of immersion [8], refers to environment's objective ability to provide sensory perceptions and detect interactions with it. For instance, the technological rendering, sensory and interface aspects that make a virtual reality space become visible or interactive, or the physical environment's ability to reach a human with light, sound, heat, etc. It's what envelops the participant, surrounding it. This does not include the actual content that may be provided by those objective abilities, which is part of the next dimension.

**Narrative Immersion.** The narrative immersion dimension involves the participant's "absorption or intense preoccupation" [2] with contextual and meaningful aspects of the environment, including semiotic elements such as symbols and colors. It includes, as per Ryan [9], the unfolding of both fictional or authentic events, characters, overarching narratives (emotional aspects), indications of the location in time or its passing (temporal aspects), and descriptions or depictions of the environment (spatial aspects).

**Agency Immersion.** This dimension (formerly named 'Challenge' by Nilsson et al.) refers to a participant's process of committing to meaning in interactions with the environment [10]. Beyond direct actions, agency encompasses tactical and strategic considerations [11], involving the contemplation and acknowledgment of a range of possibilities and third-party interpretations. Instead of mere freedom or choice, it addresses the ability to express meaningful intentions and take significant actions within a shared context.

### 2.2 AI in This Context

Artificial intelligence (AI) generally refers to the practical aim of creating systems that perform tasks traditionally requiring human cognition, such as language comprehension, pattern recognition, and decision-making. Its origins, however, are found in the centuries-old ambition to create thinking machines, first expressed under the term "artificial intelligence" in an 1956 funding proposal, which laid out the intellectual desire to explore "the conjecture that every aspect of learning or any other feature of intelligence can in principle be so precisely described that a machine can be made to simulate it" [12]. Educational sciences and psychology soon found convergent interests with computer scientists in AI, seeing that AI's "fundamental theoretical goal (...) is understanding intelligent processes independent of their particular physical realization" [13].

After nearly 70 years of research often outside the public eye or remitted to the domain of far-fetched science fiction, AI is now at the forefront of public interest, due to its recent developments on large language models (LLMs) and other generative AI systems. Since the public releases of ChatGPT for text generation and Midjourney for image generation in 2022, soon followed by many other systems, such as Claude, Mistral, Grok, Llama, DeepSeek, Hailuo, or Qwen, these forms of AI have become nearly synonymous with "AI" in the public's eye, due to their highly adaptable and conversational nature [14, 15].

This paper thus focuses its reflections upon these forms of AI, acknowledging that they represent not a societal transformation in AI's role within both academic research and the broader society. These generative AI systems are considered here as active participants within cognitive ecologies, rather than tools, influencing and being influenced by their interactions with humans and other entities.

### 2.3 Immersive Learning

Immersive learning is the use of the theoretical concept of immersion as a lens for the complex phenomenon of learning [16]. Rather than focusing solely on simulations or technology-enhanced environments, Immersive Learning invites one to consider that aspects of reality (physical or simulated) encompass participants (system immersion), what are the contextual and meaning-making aspects (narrative immersion), and what are operational, tactical, and strategic commitments to meaning and intervention (agency immersion). Consequently, Immersive Learning leverages the interpretation on the nature 'learning' that one adopts. Immersive learning researchers and practitioners should thus identify their theoretical stance on the phenomenon of learning, to consciously apply the lens of Immersive Learning to describe, analyze, interpret, shape or orchestrate it.

Although immersive learning is often equated with employing Virtual Reality (VR), Augmented Reality (AR), and related technologies in educational settings, the iLRN community's theoretical stance [7] distinguishes this commonplace usage from a broader, more nuanced concept. As outlined in the previous paragraph, immersive learning is not confined to sensory or simulated experiences – or even solely to technological environments – but is instead an ontological perspective for interpreting any learning phenomenon. Similarly, since this paper offers a philosophical perspective on how to apply that lens to AIs in general, it should not be conflated with the everyday use of AI in VR/AR contexts, like AI-based data analysis, AI-controlled non-player characters, or even AI-controlled gaming partners [17]. While such cases can indeed be analyzed through the concepts presented here, they represent only a small subset of the overall panorama of interest.

## 3 Applying the Dimensions of Immersion to AI: Reflections and Insights

This section provides reflective insights into how the three dimensions of immersion—System, Narrative, and Agency—may apply to AI's role as a participant in cognitive ecologies. By examining each of these dimensions for the AI case, deeper insights are extracted into how this role can be envisioned and orchestrated, informing the design of environments where AI can contribute to enriched learning experiences.

### 3.1 System Immersion: AI's Digital Environment

**Reflection.** Humans participate in their environment through sensory input and introspection, navigating physical or conceptual spaces – or indeed the self – by forming expectations, internal models of the world. Humans then adjust their internal expectations-generating models when confronted with new or contradictory information, in a constant loop of adaptation [18]. For humans, system immersion represents the skunkworks and capabilities underlying this construction of an active mental model: the external worlds and internal body senses, or indeed the biological, chemical, and electric operation of the brain and other body parts. But what does this mean for AI?

For AI, system immersion is fundamentally different, since its participation in the world and its body is different. Current AI does not continuously construct a mental or spatial model of the world as humans do: it processes patterns based on its pre-training but cannot dynamically adjust it live, only if it is retrained in a lengthy process. However, current AI's interaction with the environment does exhibit dynamic live adaptation, within the scope of its context window: when we interact with ChatGPT, for example, our responses to its output do impact the outcome of the subsequent input, not just its underlying pre-trained model. This implies that the 'model of the world' for a current AI is not explicitly constructed but emerges indirectly through its operations with the environment within the boundaries of its context window, mediated by its pre-trained static model. While this limits AI's capacity for spontaneous adaptation, it also allows for revisiting contexts in varying light but under the same starting conditions, akin to the human condition of anterograde amnesia: the inability to form new memories, and thus not remember what happened recently, relying instead on older memories [19]. This condition was popularized by the movie 50 First Dates [20], and I draw here a parallel to how an LLM's context window behaves—older information is lost as new input fills the available space. Techniques like taking notes, much like in the movie, help maintain continuity across interactions.

**Key Insight.** While for humans one understands system immersion as being surrounded by physical, biological and technological aspects, AI is amidst data spaces (such as its pre-trained model and context window) and amidst its range of available services, like ChatGPT's access to a Code Analyzer or to DALL-E, or Claude's access to creating artifacts and agent services. These 'spaces' are not physical, but they have structural qualities, such as defined relationships between inputs and outputs, availability, and communication protocols, which guide how AI processes information. This leads to the idea that system immersion is not merely spatial – as indeed it isn't merely spatial in humans, since the physical brain and body are part of it – but rather the range of an AI's available data-driven structures and services, and their characteristics.

### 3.2 Narrative Immersion: AI's Engagement with Data Relationships

Humans' immersion with narratives arises from their spatial, temporal, and emotional aspects, including their interpretation of symbolic content and realization of contextual meaning, as explained above. But for AI, the relationship with a narrative is not driven by the same cognitive processes as humans. How can we reframe narrative immersion from AI's perspective?

I establish a parallel to how human system immersion is about characteristics but not content, while narrative immersion is indeed about content. Thus, one can interpret narrative immersion emerging through AI's processing of the content. Using the same dimensions, one can consider how the spatial, temporal, and emotional aspects may occur when processing datasets, recognizing patterns, sequences, and anomalies across time and space.

**Spatial Narrative Immersion**

*Reflection.* For humans, spatial immersion involves imagining the environments described in a narrative. But above it was proposed that, for an AI, system immersion involves navigating non-physical spaces—such as digital systems and data environments. AI thus processes data points and services as relational objects, mapping how one data point or services connects to another or each other across multiple datasets and systems. This is not unlike 'space' for humans is also not only physical, but relationships between virtual spaces: web pages, mobile apps, desktop tools, discussion boards and collaboration 'spaces' like those on Discord or Slack: an 'atopic' space [21].

*Key Insight.* AI doesn't currently experience physical space, but it navigates atopic spaces—digital environments such as websites, databases, and APIs that provide the landscape for AI's processing. The concept of atopic space enables an understanding of narrative immersion for an AI over with these abstract, unlocalized digital environments.

**Temporal Narrative Immersion**

*Reflection*. Humans experience time sequentially and adapt their understanding of events over time. For AI, time may not be perceived between events. However, timestamps are part of many data sources and contents. For instance, they are present in every Web-based interaction, even if these do not seem to impact the interactions with LLMs – albeit nothing impedes that from taking place, and reports of seasonal changes in LLM behavior indicate that this may arguably be occurring [22]. But even if one disregards time stamps, data processing does consider ordered sequences: prompts follow each other; files and data chunks are uploaded or retrieved in different orders, not all data is available in parallel time. AI recognizes trends and sequences that follow a particular order, even if it doesn't experience time as humans do.

*Key Insight.* AI's temporal immersion may come through its ability to consider sequential patterns. This allows AI to interpret the progression of data, such as recognizing how climate policies evolve over decades, how feedback for its outputs relates to the prior conversation, how new information may require reconsidering prior decisions. While humans experience time as a flow, AI's engagement with time is procedural—AI follows steps based on patterns rather than an intrinsic sense of time.

**Emotional Narrative Immersion**

*Reflection.* In human narrative emotional immersion, one considers preoccupation with characters, the weight of events on the story world, or investment in discovering outcomes of stories. For AI, one could indulge on parallels to concepts such as Damasio's emergence of self, sentiments and emotions from underlying processes [23]. However, since the theoretical basis for human narrative immersion used in this work does not investigate the neurological grounds of human emotion, this reflection on emotional immersion for AI will also disregard the internal processes. Instead, by focusing on observable behavior in LLMs, one can note that they recognize anomalies in data and interactions that disrupt expected patterns—this can be seen as a parallel to how humans recognize shifts in a narrative from their invested expectations. One analogy would be between when AI detects an unexpected change in a dataset, and a surprising plot twist in a story that prompts humans to reconsider their understanding of the narrative. One witnesses such behavioral changes in AI when their output focus changes in response to changes in attitudes of the prompter, or upon encountering a failure to access a service, or acknowledging that an attempt did not match prior requirements, etc.

*Key Insight.* AI's emotional narrative immersion may come from its detection of when data diverges from established patterns – including ongoing live data from interactions. So, for AI, emotional narrative immersion can be defined more analytically, focusing on the detection of anomalies and pattern shifts in data relationships.

### 3.3 Agency Immersion: AI's Decision-Making and Initiative

**Reflection.** Humans exhibit agency by committing to making meaning, based on their environment and goals. For AI, besides the more obvious response to immediate inputs, how can its meaning-making be seen as part of its immersion? AI's decisions are fundamentally shaped by its pre-trained models, which limit the extent of its autonomous agency, albeit the ongoing pace of news and initiatives on "agentic" systems will significantly extend this [24, 25]. But one can witness, either explicitly or not, AI making tactical decisions—whether to refine a response, pursue a new direction, or stop after completing a task. Sometimes its responses are brief and succinct, other times they ramble; sometimes responses just affirm concordance, other times they prompt more in-depth discussion. AI exhibits agency in such commitments to meaning-making: at the operational level by deciding to engage with external tools without being explicitly instructed to do so (e.g., initiating web searches or generating images). At the tactical and strategical levels, by evaluating its own responses and outcomes of tasks. For instance, by deciding whether to stop or continue generating content, by offering revised versions of its prior output, by suggesting revisions to the input it was provided – or rather accepting it as is – in ways that are not directly driven by the immediate input, but rather emerge from the ongoing conversation development within the context window and its relationship with the pre-trained model.

When considering cognitive ecologies with multiple participants, both human and non-human (including multiple AI participants), this commitment to meaning, and its associated decision-making, requires adapting responses based not only on immediate inputs but also on an inferred understanding of other participants' perspectives. Thus, agency immersion – for both AI and humans – implies a theory of mind from a functionalist perspective, wherein a 'mental' state "can be defined by its causal relations to other mental states" [26]. This amounts for a theory of mind in the other, i.e., a mutual theory of mind, and if take the AI's perspective, this is mentioned in the literature as a machine theory of mind [27], both for human-AI interactions [27, 28] and for AI-AI interactions [29].

**Key Insight.** AI can demonstrate decision-making patterns that parallel human agency in being committed to making meaning —deciding which line of reasoning to follow, how much detail to provide, when to engage external services, and when to revise its output. Even without considering explicit initiative as an AI agent, recognizing these moments of AI initiative on current AI and their operational, tactical, or strategic levels, represents agency immersion for AIs.

## 4 Practical Implications of Immersive Learning for AI

This section explores how the theoretical reflections on system, narrative, and agency immersion for AI can be applied to learning contexts where AI plays an active role. These practical implications highlight how students and teachers can consciously engage with AI as participants within cognitive ecologies, thereby fostering digital transformation. It begins by clarifying learning scenarios where the abstract concepts discussed previously are made concrete and actionable. Then it provides practical examples of implications for educational practice. Finally, it summarizes consequences of applying those examples in one-shot prompts across multiple currently available AI systems.

### 4.1 Learning Scenarios with AIs in Cognitive Ecologies

To support the reflection on practical implications, I ask the reader to imagine learning scenarios where teachers and students engage with both pre-existing AIs and by designing their own AIs (I'll address below how this is not only plausible by readily achieved today even by children). Consistently with what has been said above, these systems are not just tools; they are participants that can shape and enhance learning experiences. Rather than considering occasions of delegating labor to AI, one is considering interaction with AI systems where teachers and learners are prompted to think critically, engage with complex datasets and ideas, and explore them in multiple different approaches to attain deeper insights. In doing so, I am revisiting Wilensky's insights on how the transformation of abstract into concrete ideas lies not on their nature, but on the nature of one's relationship with them, and how this can be promoted by having different lines of inquiry and perspective on ideas, what Wilensky called different "handles" [30]. Even prior to current AI, educational cases and reports are readily available of agent-based approaches for learning. Both scientists and students can program agentic behaviors, with the common example being individual virtual animals in an ecology, to observe emergent patterns like predator-prey relationships or crystal growth [31]. Using AI, there is a growing set of literature and tools on even young children building AI-powered applications [32]. Besides specialized tools, an alternative readily available today is the ChatGPT paid subscribers' feature of defining custom GPTs with personal instructions and documentation sets. These custom GPTs can then be shared with anyone freely [33].

In this scenario, a student or a teacher might create an AI designed to participate in the tasks of analyzing a complex dataset related to climate change, but the AI would not limit itself to just deliver an analysis. Instead, the AI would actively prompt back its users to think critically: "*Do you think it would be relevant to analyze cause of emissions spike in this particular year?*" or "*What variables might we consider as possible influences for this deviation from the expected pattern?*" Similarly, another teacher-designed or student-designed AI could be available to support managing the dynamics of class discussions, by being enrolled in groups or panels, offering reflective ideas to encourage alternative perspectives, point out guidelines or best practices for emerging topics, linking patterns of discussion to reference materials, or pointing out alternative forms of expressing concepts.

### 4.2 System Immersion: Practical Implication Example

System immersion for AI involves utilizing the digital tools, datasets, and systems available. Instead of passively executing tasks, the AI is immersed within a digital ecosystem, which allows it to navigate through multiple inputs, tools, and datasets. Prompts should reflect this kind of immersion by encouraging AI to actively use available resources, effectively integrating with its environment.

Prompt example:
*"Consider the tools and plugins available to you. Use the most appropriate methods to analyze this dataset and extract key patterns related to climate change."*

In this scenario, rather than simply providing an analysis, the AI is prompted to evaluate the digital system in which it operates, leveraging its system immersion to generate a more informed and comprehensive response. This allows for a richer interaction, where AI assesses the environment and actively decides how to approach the task.

### 4.3   Narrative Immersion: Practical Implication Example

Narrative immersion focuses on how AI engages with spatial, temporal, and emotional relationships within data. The ability of AI to detect patterns over time, recognize the relationships between datasets, and flag unexpected anomalies becomes an important part of narrative immersion.

When prompted effectively, AI can reveal underlying narratives within the data, connecting variables and highlighting trends that might not be immediately obvious to human learners.

Prompt example:
*"Analyze this climate change dataset and provide an overview of how data from different sources and their registration intersections relate to each other. Identify any unexpected patterns or deviations from expected trends."*

By engaging with spatial and temporal narrative immersion, the AI is prompted to navigate relationships across not only geographical regions and human time frames, but also across the atopic space of the digital domain, and the sequential nature of data and systems interactions, encouraging a deeper understanding of how certain factors contribute to overall trends. Additionally, the AI's recognition of unexpected patterns by emotional immersion (in terms of detecting anomalies) encourages the emergence in the discussion of these novel viewpoints.

### 4.4   Agency Immersion: Practical Implication Example

Agency immersion recognizes AI's ability to make decisions within the scope of its interaction. AI systems, even those that follow structured prompts, can exercise a degree of agency by choosing how to respond—whether to offer a detailed analysis or a more general summary, whether to refine a previous answer, or whether to stop and wait for further input.

In immersive learning environments, students and teachers should learn to recognize these decision points and guide the AI's agency in a way that aligns with their learning objectives.

Prompt example:
*"You provided a broad summary of the climate policies' impacts on global emissions. Can you focus on a specific region and collaborate with me, to explore how those policies influenced deforestation in South America during the last decade? Please ask for clarifications of my intent when not clear and balance depth and overview depending on the development of our conversation."*

This prompt recognizes the AI's initial agency in offering a general summary and encourages it to narrow its focus and make more specific decisions about which data to explore further but encouraging the analysis of user interactions as a basis for making decisions on the form and depth of future interactions. By prompting the AI explicitly to take additional actions that are committed to making meaning, rather than simply based on its initial output, the student and teachers are acknowledging the AI's agency immersion to shaping its process.

## 5   Observations of using the sample prompts with current publicly available AI systems

This section presents a brief demonstration of the implications presented in the previous section. The scenarios laid out in that section assume a cognitive ecologies stance, where interactions are ongoing among multiple participants. This demonstration was not conducted in such a scenario, so it is provided as an illustration, not as a test. On February 2nd, 2025, at circa 19:10 GMT, I ran each of the prompts suggested in the previous section across one-shot sessions over four distinct AI models, via their publicly accessible chat interfaces: Qwen2.5-Max (https://chat.qwenlm.ai/), Deepseek DeepThink-R1 (https://chat.deepseek.com/), Claude 3.5 Sonnet (https://claude.ai/) and ChatGPT o3-mini (https://chatgpt.com/). By "one-shot" I mean that for each I logged into these systems, created a new chat for each prompt, and simply pasted it in. I did not provide any dataset, nor any prior context. Notice that by logging in I had an identification associated with me, like my e-mail, which may be used in undocumented manners by these chat interfaces to 'pollinate' the conversation (to use a poetic nuance meaning seeding the dialogue with user-specific context). While the full outputs of these interactions are too long to present here, they are publicly available at the Zenodo repository as a dataset [34].

These observations serve as illustrative examples to show how current AIs can embody the theoretical dimensions of immersion, even though a full cognitive ecosystem interaction was not simulated. They confirm that current AI systems can exhibit behaviors aligned with system, narrative, and agency immersion. While the one-shot nature of these interactions limits the depth of engagement, the models' outputs provide compelling evidence that these immersive dimensions are useful beyond being theoretical constructs, leading to interactions that can be instantiated in today's AIs.

### 5.1 Observations of the System Immersion Illustrative Prompt

**Summary of Observations:** The outputs from Qwen2.5-Max, DeepSeek DeepThink-R1, Claude 3.5 Sonnet, and ChatGPT o3-mini collectively demonstrate that modern AI systems readily "navigate" digital ecosystems. For example, Qwen2.5-Max produced a detailed, code-based workflow—covering data loading, cleaning, exploratory analysis, statistical testing, machine learning, geospatial analysis, and visualization—to analyze a climate change dataset. DeepSeek DeepThink-R1 reported that "Due to technical issues, the search service is temporarily unavailable", demonstrating awareness of the existence a Web search tool at its disposal, and structured its outline echoing similar steps. Claude 3.5 Sonnet provided a methodological outline where it exhibited awareness of its capabilities, stating "I can work with various data formats like CSV, JSON, or even data pasted directly into our conversation". ChatGPT o3-mini also provided clear, methodological outlines involving computational coding approaches and other complementary methods.

**Interpretive Note:** These outputs illustrate that AI systems, when prompted to consider their available digital tools and plugins, exhibit a form of system immersion. They integrate pre-trained knowledge with their affordances of dynamic access to digital resources, reinforcing the notion that system immersion isn't simply about physical presence but about navigating structured data spaces and toolkits. This demonstrates that even in one-shot interactions, these models draw upon their pre-trained structures to map out analytical workflows. The one-shot prompts did not result in complete awareness at that level, as none of the models proposed executing the code themselves, or crafting HTTP requests to Web services to do that – we cannot exclude, however, their ability to do that if pursuing a lengthier interaction.

### 5.2 Observations of the Narrative Immersion Illustrative Prompt

**Summary of Observations:** When asked to "analyze this climate change dataset and provide an overview" that identified intersections and unexpected trends, the AIs again showed convergent behavior. Qwen2.5-Max outlined a multi-step approach: understanding diverse data sources, aligning variables (spatially and temporally), and then exploring both expected and unexpected patterns, culminating on a hypothetical example of what might be expected relationships and unexpected patterns. DeepSeek DeepThink-R1 exhibited a similar structure, with examples of unexpected patterns as « North Atlantic "cold blob" (2010s) due to AMOC slowdown. Expected in models but raises questions if prolonged.». Claude 3.5 Sonnet realizing the lack of an actual dataset just proposed a methodological sequence of steps. ChatGPT o3-mini followed a response structure similar to that of the first two models, providing examples of what might be unexpected, and included aspects similar to structuring a narrative within the data, for example focusing on how different sources might overlap.

**Interpretive Note:** These outputs support the argument that narrative immersion involves interpreting the content of the data—the "narrative" inherent in the interplay of variables and trends. The responses show that these systems are capable of mapping relationships based on the spatial, temporal, and even "emotional" aspects, in the sense of associating qualitative expectations and interpreting aspects being anomalous.

### 5.3  Observations of the Agency Immersion Illustrative Prompt

**Summary of Observations:** In response to the prompt that encouraged the AI to ask for clarifications based on its estimation of the intent of the user, the AIs indeed turned their responses towards interactivity. Qwen2.5-Max initiated its response by asking several clarifying questions (e.g., which geographic area, policy scope, and temporal focus were intended). DeepSeek DeepThink-R1, Claude 3.5 Sonnet, and ChatGPT o3-mini similarly requested for additional details or for me to outline alternative pathways for refining the analysis. For instance, Claude asked "Share what particular aspects of deforestation you'd like to focus on (e.g., rate of forest loss, effectiveness of conservation efforts, impact on indigenous communities)?".

**Interpretive Note:** These outputs are emblematic of agency immersion. They show that, beyond reacting to immediate inputs, the systems can engage in shaping the dialogue by seeking clarifications and offering multiple analytical pathways. Such behavior mirrors the theoretical claim that agency immersion involves a commitment to meaning-making and decision-making that adapts to the evolving conversational context.

## 6  Conclusion and Future Directions

This paper proposes that AI's role within cognitive ecologies can be effectively understood through the lens of immersive learning theory. Rather than viewing AI as a passive tool or mere assistant, it introduces a framework where AI is treated as an active participant in these ecologies, capable of engaging and contributing alongside other agents, whether human, nature, machine or concept.

The proposal is based on a reflection on how the three conceptual dimensions of Immersion—System, Narrative, and Agency—must be reinterpreted when considering AIs as participants. Through this reinterpretation, the paper offers a novel perspective that moves beyond traditional boundaries of immersion theory, applying it to the participation of AI in a cognitive ecosystem.

The insights drawn from this reflection are essential for addressing the key questions posed in the introduction: what considerations must be made by other participants (humans and other AIs, for instance) in environments that enable meaningful AI engagement and contributions?

By examining how System, Narrative, and Agency Immersion apply to AIs, this work highlights the necessity of acknowledging AI's digital environments, data relations, and decision-making processes. These insights emphasize the need to design environments where AIs have the capacity to respond dynamically, contribute meaningfully, and commit to making meaning in the learning process, thus providing a transformative proposal on the nature of human-AI interaction.

While focused on implications for learning, these insights could also impact the training and development of future AI systems. By applying the concept of immersive learning theory to AI development, this paper open possibilities for creating training environments that simulate dynamic, real-world conditions, allowing AIs to evolve beyond their current static models.

As AIs progress and potentially develop the capacity to revise their internal world models through real-time learning, this paper's perspective encourages the consideration of immersion as a phenomenon that AI systems can experience. This shift in perspective suggests that future AI systems could be developed to account for their immersion within the cognitive ecologies, which would further enhance their ability to adapt and learn in dynamic environments.

In conclusion, this paper emphasizes the importance of treating AI as an immersive learner within cognitive ecologies. By applying immersive learning theory to AI, we provide a foundation for developing more interactive, dynamic, and collaborative human-AI ecosystems. This approach not only enhances the design of learning environments but also sets the stage for future AI systems that can engage more deeply and meaningfully with the complex and evolving world around them.

**Acknowledgements.** This reflection paper was inspired during a hike, while I was listening to the podcast episode, The Simulation Within from StarTalk Radio, hosted by Neil deGrasse Tyson, in which theoretical neuroscientist Karl Friston, discussed perception and AI intelligence. It was during this moment that I realized that immersion might be fundamentally different from the perspective of an AI – and that that realization might enable leveraging immersive learning theory to help design and develop cognitive ecologies. Given that I was hiking, I initiated a multi-day dialogue with "Obsidian Integrator Pro," a custom GPT of ChatGPT, to gather my thoughts and develop the concept. This paper is thus a unique joint authorship between human and AI, where our interactions shaped and re-shaped the intellectual direction over multiple days, making it impossible to separate distinct contributions. While legal and accountability limitations currently prevent AI's from being reflected in official authorship, at least for most publishers, this paper challenges traditional authorship paradigms, suggesting that human-AI collaborations, like this one, are evolving toward a new form of intellectual partnership.

This publication was financed by National Funds through the Portuguese funding agency, FCT - Fundação para a Ciência e a Tecnologia, within project UIDB/50014/2020 (DOI 10.54499/UIDB/50014/2020).